\batchmode
\makeatletter
\def\input@path{{C:/Arbeiten/}}
\makeatother
\documentclass[english]{article}
\usepackage[T1]{fontenc}
\usepackage[latin9]{inputenc}
\usepackage{color}
\usepackage{amsmath}
\usepackage{amssymb}
\usepackage{babel}
\begin{document}
\title{Distributional Metrics and the Action Principle of Einstein-Hilbert
Gravity}
\author{Albert Huber\thanks{ahuber@tph.itp.tuwien.ac.at}}
\date{{\footnotesize{}Institut für theoretische Physik, Technische Universität
Wien, Wiedner Hauptstr. 8-10, A-1040 Wien, AUSTRIA}}
\maketitle
\begin{abstract}
In this work, a subclass of the generalized Kerr-Schild class of spacetimes
is specified, with respect to which the Ricci tensor (regardless of
the position of indices) proves to be linear in the so-called profile
function of the geometry. Considering Colombeau's nonlinear theory
of generalized functions, this result is extended to apply to an associated
class of distributional Kerr-Schild geometries, and then used to formulate
a variational principle for these singular spacetimes. More specifically,
it is shown in this regard that a variation of a suitably regularized
Einstein-Hilbert action can be performed even if the metric of one
of the corresponding generalized Kerr-Schild representatives contains
a generalized delta function that converges in a suitable limit to
a delta distribution.
\end{abstract}
\textit{\footnotesize{}Key words: generalized Kerr-Schild class, Colombeau
algebra, action principle}{\footnotesize\par}

\section*{Introduction}

\textcolor{black}{The general theory of relativity is a nonlinear
theory of gravity. The mathematical theory of distributions, on the
other hand, is a linear theory that uses a variety of techniques which
cannot be implemented in the nonlinear framework of Lorentzian geometry.
This is also true with regard to Colombeau's nonlinear theory of generalized
functions \cite{colombeau2000new,colombeau2011elementary,grosser2001geometric,grosser2002global,kunzinger2002foundations},
which, though capable of solving an impressive spectrum of problems
associated with the treatment of distributions in gravitational physics
does not always allow a rigorous treatment of the simultaneously singular
and nonlinear field equations of theory.}

In view of this fact, it is surprising that in Einstein's theory well-defined
distributional metrics and curvature expressions have been found in
the past. These offer the possibility of characterizing singular energy-momentum
distributions and thus can be used to solve the highly non-trivial
problem of how to describe gravitational fields of point-like particles
in general relativity.

Perhaps the most prominent classes of solutions of Einstein's equations
characterizing the gravitational fields of said point-like sources
are, on the one hand, the class of impulsive pp-wave spacetimes and,
on the other hand, the class of gravitational shock wave spacetimes
in black hole and cosmological backgrounds. The most famous representatives
of the first class are the Aichelburg-Sexl geometry \cite{aichelburg1971gravitational}
and the models of Lousto and Sanchez \cite{lousto1989gravitational},
which arose as ultrarelativistic limits of boosted black hole spacetimes
of Schwarzschild and Reissner-Nordström geometries. The most famous
representatives of the second class, on the other hand, are the shock
wave geometries found by Dray and 't Hooft \cite{dray1985gravitational}
and Sfetsos \cite{sfetsos1995gravitational}, which, as recently discovered,
can be obtained as a special case of a more general class of gravitational
shock wave geometries in stationary black hole backgrounds \cite{Huber2019ultpar}.

The representatives of both of these classes have in common that they
have a delta-like profile, which implies that their geometries are
regular everywhere, with the exception of a special null hypersurface
on which the confined field of the ultrarelativistic particle is concentrated.
Unfortunately, due to the nonlinear structure of the field equations
of the theory, said representatives also have in common that they
lead to very problematic curvature expressions, which are pathological
insofar as they contain ill-defined products of delta distributions. 

More specifically, ill-defined 'squares of the delta distribution'
have occurred in the past in the course of the calculation of gravitational
shock wave spacetimes in black hole backgrounds as a result of the
(rather careless) application of Penrose's 'scissors-and-paste' method
\cite{penrose1972geometry}. This left the authors of the cited works
with no choice but to 'blithely ignore' all problematic terms and
solve only the remaining meaningful part of Einstein's field equations. 

This, of course, did not solve the problem. Fortunately, however,
a few years later it became apparent that the generalized Kerr-Schild
framework can be used to show that (in a specific sense) the results
obtained are meaningful not only from a physical, but also from a
mathematical point of view \cite{alonso1987generalized,balasin2000generalized}.
The main reason for this is that the said geometric framework has
the special property that the mixed deformed Einstein tensor is linear
in the so-called profile function of geometry; a circumstance that
can be used to both determine the geometric structure of the metric
of spacetime and to avoid the dangerous ill-defined terms arising
in the course of the calculation. This trick was then also used to
determine the exact structure of the shock wave geometries calculated
in \cite{Huber2019ultpar}.

However, the problem with this idea of treating the subject is that
it still proves to be impossible to deduce the corresponding distributional
geometries from Einstein's equations with lowered and raised indices.
In addition, it turns out to be impossible to obtain the field equations
by varying the Einstein-Hilbert action, so that one could get the
impression that there are specific models in general relativity that
cannot be deduced from the principle of stationary action. 

In response to this particular shortcoming of the theory, it shall
be proven in the following that both the class of impulsive pp-wave
spacetimes as well as the class of gravitational shock wave spacetimes
are contained in the superordinate class of spacetimes, which represents
a specific subclass of the generalized Kerr-Schild class, to be discussed
in section one of this work. The representatives of this class share
the main geometric properties that their associated Kerr-Schild null
vector fields are the generators of a foliation of null hypersurfaces
and that the corresponding profile functions remain constant under
the flow of these Kerr-Schild generators. This has the effect that
not only the Ricci tensor with mixed indices, but also that with lowered
and raised indices are linear in the profile function of the geometry.
Based on these findings, it is concluded that not only can corresponding
linear field equations be formulated, but also that it should be possible
to perform a variation of the Einstein-Hilbert action with a degree
of mathematical rigor sufficiently high to obtain a meaningful result
in the end. 

The strategy to perform said variation, which is explored in section
two of this work, is to use methods from the theory of Colombeau algebras
of manifold-valued generalized functions; a mathematical theory tailor-made
for dealing with singular problems in various geometrical settings
used in gravitational physics. More precisely, the main idea in this
context is not to consider a delta distribution directly, but to consider
instead a so-called strict delta net, which converges to said distribution
in a suitable limit. The consideration of such a delta net proves
both beneficial and necessary in this regard owing to the fact that
it allows one to derive mathematically well-defined distributional
field equations whose distributional character emerges (in the limit
previously mentioned) only after the variation of the regularized
Einstein-Hilbert action has been completed. The avoidance of encountering
any ill-defined 'delta-square' terms is thereby achieved via focussing
exclusively on generalized Kerr-Schild spacetimes with linear deformed
Ricci tensor. 

The physical significance of the approach is illustrated using special,
selected examples of distributional spacetimes in the third and final
section of this work. The focus is placed, however, exclusively on
impulsive pp-wave spacetimes and gravitational shock wave spacetimes
in black hole and cosmological backgrounds. 

\section{Generalized Kerr-Schild Framework}

All spacetimes to be considered in this work are generalized Kerr-Schild
spacetimes belonging to the so-called generalized Kerr-Schild class.
Spacetimes lying in this class have the property that their metrics
can be decomposed in the form
\begin{equation}
\tilde{g}_{ab}=g_{ab}+fl_{a}l_{b},
\end{equation}
where $g_{ab}$ represents the so-called background or seed metric,
$f$ represents the so-called profile function and $l_{a}=g_{ab}l^{b}$
the so-called Kerr-Schild null co-vector field. An important factor
in this decomposition is that the associated Kerr-Schild null vector
field $l^{a}$ is given in such a way that the null geometric constraints
$\tilde{g}_{ab}l^{a}l^{b}=g_{ab}l^{a}l^{b}=0$ and $(l\tilde{\nabla})l^{a}=(l\nabla)l^{a}=0$
are met, so that a congruence of null geodesics is formed whose associated
vector field $l^{a}$ is to be assumed as affinely parametrized for
simplicity's sake.

The main property of any generalized Kerr-Schild class spacetime is
that its mixed Ricci tensor

\begin{align}
\tilde{R}_{\;b}^{a} & =R_{\;b}^{a}-\frac{1}{2}fR_{\;c}^{a}l^{c}l_{b}-\frac{1}{2}fR_{\;b}^{c}l_{c}l^{a}+\\
 & +\frac{1}{2}(\nabla_{c}\nabla^{a}(fl^{c}l_{b})+\nabla^{c}\nabla_{b}(fl_{c}l^{a})-\nabla_{c}\nabla^{c}(fl^{a}l_{b})),\nonumber 
\end{align}
its Ricci scalar
\begin{equation}
\tilde{R}=R-fR_{\;c}^{d}l_{d}l^{c}+\nabla_{d}\nabla^{c}(fl^{d}l_{c})
\end{equation}
and therefore its mixed Einstein tensor
\begin{equation}
\tilde{G}_{\;b}^{a}=\tilde{R}_{\;b}^{a}-\frac{1}{2}\delta{}_{\;b}^{a}\tilde{R}=G_{\;b}^{a}+\rho{}_{\;b}^{a}
\end{equation}
with
\begin{align}
\rho{}_{\;b}^{a} & =-\frac{1}{2}fR_{\;c}^{a}l^{c}l_{b}-\frac{1}{2}fR_{\;b}^{c}l_{c}l^{a}+\frac{1}{2}\delta{}_{\;b}^{a}(fR_{\;c}^{d}l_{d}l^{c}-\nabla_{d}\nabla^{c}(fl^{d}l_{c}))+\\
 & +\frac{1}{2}(\nabla_{c}\nabla^{a}(fl^{c}l_{b})+\nabla^{c}\nabla_{b}(fl_{c}l^{a})-\nabla_{c}\nabla^{c}(fl^{a}l_{b}))\nonumber 
\end{align}
are all linear in the profile function $f$. Obviously, this particular
property of the generalized Kerr-Schild framework proves to be extremely
useful in practice for finding exact solutions to Einstein's equations.

Obviously, decomposition relation $(1)$ allows one to associate two
different metrical structures $\tilde{g}_{ab}$ and $g_{ab}$ with
a given manifold $\tilde{M}$. The situation is similar with the inverse
metrics, which are related with each other by means of the decomposition
relation
\begin{equation}
\tilde{g}^{ab}=g^{ab}-fl^{a}l^{b}.
\end{equation}
However, the Ricci and Einstein tensors with raised and lowered indices
unfortunately do not have the same remarkable properties. Instead,
they usually turn out to be nonlinear in the profile function $f$.
Of course, the same holds true for the Riemann curvature tensor of
the geometry. 

Nevertheless, it would be desirable to know under which exact circumstances
at least the Ricci with lowered and raised indices is linear in $f$.
To investigate this, one may use relations $(1)$ and $(6)$ to set
up the affine connection 
\begin{equation}
C_{\:bc}^{a}=\frac{1}{2}\nabla_{b}(fl^{a}l_{c})+\frac{1}{2}\nabla_{c}(fl^{a}l_{b})-\frac{1}{2}\nabla^{a}(fl_{b}l_{c})+\frac{1}{2}fDfl^{a}l_{b}l_{c},
\end{equation}
which relates the pair of covariant derivative operators $\tilde{\nabla}_{a}$
associated with $\tilde{g}_{ab}$ and $\nabla_{a}$ and associated
with $g_{ab}$. Given this definition, the deformed Riemann curvature
tensor of the geometry can be set up in the next step, which has the
form

\begin{equation}
\tilde{R}_{\;bcd}^{a}=R_{\;bcd}^{a}+E_{\;bcd}^{a},
\end{equation}
provided that the abbreviation $E_{\,bcd}^{a}=2\nabla_{[c}C_{\,d]b}^{a}+2C_{e[c}^{a}C_{d]b}^{e}$
is used in the present context. Using the fact that $C_{\:ab}^{b}=0$
holds true with respect to any generalized Kerr-Schild class metric,
the deformed Ricci tensor with lowered indices reads
\begin{equation}
\tilde{R}_{ab}=R_{ab}+E_{ab},
\end{equation}
where $E_{ab}=\nabla_{c}C_{\:ab}^{c}+C_{\:ad}^{c}C_{\:cb}^{d}$ applies. 

To ensure that this object as well as the Einstein tensor of the geometry
are linear in the profile function $f$, it must be ensured that the
conditions 
\begin{equation}
C_{\:ad}^{c}C_{\:cb}^{d}\overset{!}{=}0
\end{equation}
and 
\begin{equation}
\nabla_{c}C_{\:ab}^{c}l^{a}\overset{!}{=}\nabla_{c}C_{\:ab}^{c}l^{b}\overset{!}{=}0
\end{equation}
as well as

\begin{equation}
\tilde{R}_{\;b}^{a}l_{a}l^{b}\overset{!}{=}R_{\;b}^{a}l_{a}l^{b}\overset{!}{=}0
\end{equation}
are met, where the conditions $(11)$ and $(12)$ result from the
consistency condition $\tilde{R}_{ab}\overset{!}{=}\tilde{g}_{ac}\tilde{R}_{\;b}^{c}$.
However, using the result

\begin{equation}
C_{\:ad}^{c}C_{\:cb}^{d}=\frac{1}{2}\left\{ [(l\nabla)f]^{2}+f^{2}\nabla^{[c}l^{d]}\nabla_{d}l_{c}+f^{2}\nabla_{[d}l_{c]}\nabla^{c}l^{d}\right\} l_{a}l_{b},
\end{equation}
it immediately becomes clear that both the Ricci and the Einstein
tensors with lowered indices are linear in the profile function if
the conditions 
\begin{equation}
\tilde{\nabla}_{[a}l_{b]}=\nabla_{[a}l_{b]}\overset{!}{=}0
\end{equation}
and

\begin{equation}
L_{l}f=(l\nabla)f\overset{!}{=}0
\end{equation}
are met, where $L_{l}$ is the Lie derivative with respect to $l^{a}$.
It must therefore be assumed that the profile function of the geometry
can be selected exactly in such a way that it vanishes along the flow
of the vector field $l^{a}$. Furthermore, said vector field must
be chosen exactly in such a way that it represents the generator of
a foliation of spacetime in lightlike hypersurfaces, generally referred
to as null foliation of spacetime\footnote{Note that a large number of different constructions of said null foliations
has been given in the literature over the years \cite{friedrich1999rigidity,hayward1993dual,moncrief1983symmetries},
some of which are based on very different mathematical and physical
assumptions. A quite recent construction, which is strongly based
on previous results on so-called double null foliations of spacetime,
is discussed in \cite{Huber2019foliation}.}. It is therefore clear that the class of spacetimes to be considered
is a specific subclass of the Robinson-Trautmann class of spacetimes,
which is a class of spacetimes, admitting a geodesic, shearfree, twistfree
and diverging congruence of null curves. 

Assuming now that this is the case, one finds that the nonlinear part
of the deformation of the Riemann tensor of the geometry is given
by the expression 
\begin{equation}
C_{e[c}^{a}C_{d]b}^{e}=2f\nabla^{e}fl_{[c}\nabla_{d]}l_{e}l^{a}l_{b},
\end{equation}
whereas the deformed Ricci takes the form 
\begin{equation}
\tilde{R}_{ab}=R_{ab}+\frac{1}{2}\nabla_{c}\nabla_{a}(fl^{c}l_{b})+\frac{1}{2}\nabla_{c}\nabla_{b}(fl^{c}l_{a})-\frac{1}{2}\nabla^{2}(fl_{a}l_{b}).
\end{equation}
Hence, looking at relation $(16)$, it can be concluded that $C_{e[c}^{a}C_{d]b}^{e}=0$
is met if and only if there exists a null frame $(l^{a},k^{a},m^{a},\bar{m}^{a})$
such that $(m\nabla)l^{a}\propto l^{a}$ applies and there is a profile
function $f$ in relation to which not only $(l\nabla)f=0$, but also
$(l\nabla)(m\nabla)f=(m\nabla)(l\nabla)f=0$ is fulfilled. This is
because in such a case 

\begin{equation}
\nabla^{e}fl_{[c}\nabla_{d]}l_{e}=l_{[d}(l\nabla)\nabla_{c]}f=0
\end{equation}
applies, which, however, implies that the deformed Riemann tensor
must be of the form 
\begin{equation}
\tilde{R}_{\;bcd}^{a}=R_{\;bcd}^{a}+2\nabla_{[c}C_{\,d]b}^{a}.
\end{equation}
This result represents the very last needed for further investigations.

As shall be shown in the following, the results obtained are of great
importance for the derivability of special classes of distributional
solutions of Einstein's field equations from a suitably regularized
Einstein-Hilbert action, since they allow one to bypass the problem
of performing nonlinear operations on distributional objects. The
definition of said regularized action functional is thereby achieved
in this context on the basis of well-established methods of Colombeau's
theory of generalized functions, which allow one to avoid ill-defined
products of distributions and to derive well-defined Euler-Lagrange
equations from a variation of the regularized Lagrangian of the theory.
This shall be explained in more detail below.

\section{Colombeau Algebras and the Einstein-Hilbert Action}

Based on the results obtained in the previous section, one may now
proceed by considering the special case of a generalized Kerr-Schild
spacetime with smooth background metric and a profile distribution
containing a delta-like singularity being concentrated on a single
null hypersurface of spacetime. More precisely, given some double
null coordinate system $(u,v,\theta,\phi)$, one may consider the
case of a Kerr-Schild metric $(1)$ with profile distribution $f=f(u,v,\theta,\phi)$
of the form
\begin{equation}
f=f_{0}\delta,
\end{equation}
where $\delta\equiv\delta(u)$ is Dirac's delta distribution and $f_{0}=f_{0}(v,\theta,\phi)$
is a function of the remaining coordinates, from now on to be referred
to as reduced profile function. It then turns out that $(k\nabla)f=f_{0}(k\nabla)\delta$
must be valid if $(k\nabla)u=1$, where $k^{a}$ is the null vector
field non-tangential to the $u=0$-null hypersurface $\mathcal{X}$.

Without referring to a concrete geometric model at this point, it
shall however be assumed for the sake of simplicity that there is
a normalized null frame $(l^{a},k^{a},m^{a},\bar{m}^{a})$, which
can be chosen in such a way that not only conditions $(12)$, $(14)$
and $(15)$, but also the conditions $(m\nabla)l^{a}\propto l^{a}$
and $(l\nabla)(m\nabla)f=(m\nabla)(l\nabla)f=0$ are met; although
it is completely sufficient if this applies only locally at $\mathcal{X}$. 

Keeping in mind that the calculation of the curvature tensor and its
invariants requires to perform nonlinear operations involving both
the metric and its inverse, both of which contain a Kerr-Schild deformation
proportional to a delta distribution, great care must be taken at
this point; especially in view of the fact that the definition of
the Einstein-Hilbert action for a generalized Kerr-Schild spacetime
with profile function of the form $(20)$ requires the calculation
of the square root of the determinant of the Kerr-Schild metric.

Consequently, to avoid serious mathematical deficiencies related to
the consideration of a delta distribution in the Kerr-Schild metric
$(1)$, the strategy of approaching the problem of how to derive distributional
field equations from the Einstein-Hilbert action in the face of the
low regularity of the profile function will be to resort to Colombeau's
theory of generalized functions \cite{colombeau2000new,colombeau2011elementary,grosser2001foundations,grosser2002global,kunzinger2002foundations,kunzinger2002generalized,kunzinger2003intrinsic,kunzinger2009sheaves}.
The basic idea behind this strategy is that Colombeau's theory provides
a suitable framework for a mathematically rigorous treatment of problems
associated with the differentiation and execution of nonlinear operations
on singular quantities that actually arise as products of distributions
on either Riemannian or Lorentzian manifolds. 

For the reader not so familiar with said theory, a brief introduction
will now be given; although only the most important facts and pieces
of non-redundant information will be covered in the following. To
get a better overview of the theory, of course, it is advisable to
consult more detailed and mathematically precise treatments of the
subject, such as, for example, given in \cite{grosser2001geometric}.

Since it provides a flexible and efficient way of modelling singularities
in general relativity, the focus of the introduction shall be placed
right away on the so-called special (or simplified) Colombeau framework,
dealing with so-called special Colombeau algebras \footnote{As shall be clarified below, \textit{special} Colombeau algebras are
to be distinguished from so-called \textit{full} Colombeau algebras.} of manifold-valued generalized functions. Given a paracompact $C^{\infty}$-manifold
$X$, the center of attention of this special (or simplified) Colombeau
framework is the so-called special Colombeau algebra $\mathcal{G}(X)$,
which is a commutative, associative and unital differential algebra
that contains the vector space of Schwartz distributions as a linear
subspace, and the space of smooth functions as a faithful subalgebra.
As such, it is an algebra consisting of one-parameter families of
$C^{\infty}$-functions $(f_{\varepsilon}(x))_{\varepsilon\in(0,1]}$,
which are subject to certain growth conditions in $\varepsilon$.
To be more precise, $\mathcal{G}(X)$ results from forming the quotient
algebra $\mathcal{E}_{m}(X)/\mathcal{N}(X)$ of the algebra of nets
of moderate functions $\mathcal{E}_{m}(X)=\{(f_{\varepsilon})_{\varepsilon}\in C^{\infty}(X)^{(0,1]}:\forall K\subset\subset X\:\forall P\in\mathcal{P}(M)\:\exists l\:\underset{x\in K}{\sup}\vert Pf_{\varepsilon}(x)\vert=O(\varepsilon^{-l})\}$
by the ideal of nets of so-called negligible functions $\mathcal{N}(X)=\{(f_{\varepsilon})_{\varepsilon}\in C^{\infty}(X)^{(0,1]}:\forall K\subset\subset X\:\forall m\:\forall P\in\mathcal{P}(M)\:\underset{x\in K}{\sup}\vert Pf_{\varepsilon}(x)\vert=O(\varepsilon^{m})\}$,
where, in this context, $\mathcal{P}(X)$ denotes the space of all
linear differential operators on the manifold $X$. 

Probably the most significant advantage of working with the Colombeau
framework is that it extends the standard repertoire of operations
available for theories of distributions and smooth functions, respectively.
In particular, said mathematical framework can be used not only in
a strictly linear, but also in a nonlinear context, where conventional
linear distribution theory has its natural limitations \cite{schwartz1954limpossibilite}.
The main reason for this is that the corresponding algebras of generalized
functions yield expressions that are singular in a fixed, but in principle
arbitrary real (regularization) parameter $\varepsilon$, which coincide
with Schwartz distributions in the limit $\varepsilon\rightarrow0$
(if such a limit exists). In this sense, said elements of $\mathcal{G}(X)$
can be identified as regularizations of distributions, which, as has
long been known in theoretical physics, is very satisfying in the
sense that regularizations of distributions are much easier to handle
in practice than the distributions with which they are associated.

Unfortunately, however, the special algebra $\mathcal{G}(X)$ usually
suffers from the disadvantage that its elements do not allow one to
make a single, unique choice for the parameter $\varepsilon$. This
general absence of a preferred regularization parameter is accompanied
by the lack of a preffered regularization method (in the sense that
there is no canonical embedding of distributions into $\mathcal{G}(X)$),
which in turn is the reason why there cannot be such a thing as a
single canonical Colombeau algebra. Rather, the situation is such
that there are many different types of Colombeau algebras, whose constructions
revolve around the same principles, but often are only loosely related
to each other, which has the disadvantage that the results obtained
for one Colombeau algebra often cannot be formally transferred to
another. Nevertheless, there are repeated situations in which certain
mathematical or physical assumptions can be made on the basis of which
a preferred construction can be selected.

A well-known situation, in which such a preferred choice can actually
be made, is given, for example, if $X\subseteq\mathbb{R}^{n}$ open.
The standard procedure to embed both continuous functions and compactly
supported distributions into $\mathcal{G}(X)$ in such a case is to
consider an appropriate mollifier $\rho$ satisfying

\begin{align*}
i) & \int\rho(x)d^{n}x=1,\:and\\
ii) & \int\rho(x)x^{\alpha}d^{n}x=0\;\forall\vert\alpha\vert\geq0,
\end{align*}
which can be used to set $\rho_{\varepsilon}(x)\equiv\varepsilon^{-n}\rho\left(\frac{x}{\varepsilon}\right)$. 

Given this choice, the embedding of elements of $\mathcal{D}'(X)$
- the space of Schwartz distributions on $X$ - into $\mathcal{G}(X)$
is then accomplished by considering generalized functions $f_{\varepsilon}(x)=\int\rho_{\varepsilon}(x-y)f(y)d^{n}y$
that converge to distributions in the limit $\varepsilon\rightarrow0$.
In case that $supp(\rho)$ is not compact, a sheaf-theoretic construction
(which is based on the observation that the functor $X\rightarrow\mathcal{G}(X)$
defines a fine sheaf of differential algebras (in the category of
complex vector spaces)) must be used to fix the regularization and
to choose a specific preferred mollifier $\rho(x)$ for the convolution,
which is based on the consideration of a partition of unity subordinate
to the charts of some atlas. The embedding $\mathcal{D}'(X)\hookrightarrow\mathcal{G}(X)$
is then called canonical and $\mathcal{G}(X)$ is no longer referred
to as special, but as full Colombeau algebra.

In more general cases, however, in which $M$ is not strictly presupposed
to be an open subset of $\mathbb{R}^{n}$, the situation is more involved.
This is not least because in these more general cases the procedure
of embedding $\mathcal{D}'(X)$ into $\mathcal{G}(X)$ via convolution
with a preferred mollifier (an idea that represents one of the main
building blocks of the theory in $\mathbb{R}^{n}$) is usually prevented
by diffeomorphism invariance. Respecively, to put it more accurately,
the problem arises that Colombeau algebras which allow for a canonical
embedding of distributions generally lack the feature of diffeomorphism
invariance. 

A way out of this dilemma is to exploit the fact that de Rham regularizations
are available through convolution with a mollifier in charts. However,
the use of this finding entails the disadvantage that in order to
obtain a covariant result it is necessary to explicitly check the
coordinate invariance of the results obtained on a case-by-case basis.
For this reason, it seems more natural and elegant to consider full
diffeomorphism invariant Colombeau algebras instead, which entail
a canonical embedding of distributions and yet allow for covariant
regularization procedures in the modeling of singularities. Such algebras
have indeed been discovered a while ago and studied intensively in
the literature ever since \cite{grosser2002global,nigsch2015new,nigsch2016full,steinbauer2010geometric}.
An introduction to the technical machinery associated with the existence
of such algebras shall however be avoided at this stage in favour
of the consideration of special Colombeau algebras of manifold-valued
generalized functions. 

The main reason for this is that special Colombeau algebras, although
they lack a canonical embedding of the space of distributions, not
only allow one to model singularities in a nonlinear context broadly
and efficiently, but also lend themselves in a very natural way to
geometric applications. As a result, they offer a suitable framework
in any situation in which one is prepared to refrain from such a canonical
embedding, that is, in particular, when considering models that are
given in relation to a fixed coordinate system. Such models will be
the subject of this work in due course.

The interplay between generalized functions and distributions is most
conveniently formalized in terms of the notion of weak equality or
association. A generalized function $(f_{\varepsilon}(x))_{\varepsilon}$
and a distribution $T$ are called locally associated if 
\begin{equation}
\underset{\varepsilon\rightarrow0}{\lim}\int f_{\varepsilon}\nu\equiv\underset{\varepsilon\rightarrow0}{\lim}\langle f_{\varepsilon},\nu\rangle\equiv\langle T,\nu\rangle
\end{equation}
for all compactly supported one-densities $\nu$ on $X$. In such
a case, one writes $f\thickapprox T$. On the other hand, two generalized
functions $(f_{\varepsilon}(x))_{\varepsilon}$ and $(g_{\varepsilon}(x))_{\varepsilon}$
are associated if $f-g\thickapprox0$. Hence, as can readily be seen,
association behaves like equality on the level of distributions. It
is an equivalence relation compatible with addition and differentiation
and it allows multiplication with $C^{\infty}$ functions. However,
as is well known, it does not respect multiplication of Colombeau
objects.

The simplest way to illustrate this is to consider classic examples
in $\mathbb{R}^{n}$; a case, in which the association relation $f\thickapprox T$
for a generalized function $(f_{\varepsilon}(x))_{\varepsilon}$ and
a distribution $T$ formulated in $(21)$ gives the expression

\begin{equation}
\underset{\varepsilon\rightarrow0}{\lim}\langle f_{\varepsilon},\varphi\rangle\equiv\langle T,\varphi\rangle,
\end{equation}
which can be written down somewhat less compactly in the form

\begin{equation}
\underset{\varepsilon\rightarrow0}{\lim}\int f_{\varepsilon}(x)\varphi(x)d^{n}x\equiv\int T(x)\varphi(x)d^{n}x
\end{equation}
for all $\varphi(x)\in C_{0}^{\infty}$. Perhaps the most well-known
example occurs if one tries to calculate the powers of the $\theta(x)$
function, which upon naive multiplication would lead to serious contradictions.
Specifically, if one tries to conclude 
\begin{equation}
\theta^{n}(x)=\theta(x)\Rightarrow n\theta(x)^{n-1}\theta'(x)=\theta'(x),
\end{equation}
one immediately finds that this cannot hold for arbitrary $n$, since
it would imply the validity of 
\begin{equation}
(\theta(x)^{n})'=n\theta(x)\theta'(x)=\theta'(x)=(\theta(x)^{n+1})'=(n+1)\theta(x)\theta'(x)
\end{equation}
and thus would force one to erroneously conclude that $\theta'(x)=0$.
However, since one would also expect that $\theta'(x)=\delta(x)$,
this would also imply that $\delta(x)=0$, which is obviously nonsense.

Of course, from the point of view of Colombeau theory, the situation
is different. There, one rather has
\begin{equation}
\theta^{n}(x)\thickapprox\theta(x)\Rightarrow n\theta(x)^{n-1}\theta'(x)\thickapprox\theta'(x),
\end{equation}
which, after using the fact that $\theta'(x)\thickapprox\delta(x)$,
where $\delta(x)$ is Dirac's delta distribution, leads to the results

\begin{equation}
\theta(x)\cdot\theta'(x)\thickapprox\frac{1}{2}\delta(x)
\end{equation}
and
\begin{equation}
\theta^{n}(x)\cdot\theta'(x)\thickapprox\theta(x)\cdot\theta'(x)\thickapprox\frac{1}{n+1}\delta(x).
\end{equation}
Hence, one finds that

\begin{equation}
\theta(x)\cdot\theta'(x)\thickapprox\theta(x)\cdot\delta(x)\thickapprox A\delta(x)
\end{equation}
for some constant $A$, which makes it perfectly clear that it would
be both wrong and misleading to naively conclude that in the Colombeau
algebra $\theta$ times $\delta$ is just $\frac{1}{2}\delta$. Instead,
as it turns out, association enables one to model $\theta$ times
$\delta$ in a large number of ways, which shows that the situation
is much more nuanced and the problem is much more diverse than one
would expect at first glance.

Anyway, after this brief introduction to Colombeau theory, it may
be the time to return to the main subject of this work, which is the
problem of formulating a well-defined variational principle for distributional
Kerr-Schild metrics. 

For the purpose of dealing with this subject, the first step will
be to consider a singular generalized Kerr-Schild spacetime $(\tilde{M},\tilde{g})$
with properties very similar to those of the class of generalized
Kerr-Schild spacetimes presented in section one, whereas the main
difference shall be that the profile function of the geometry is assumed
to be of the form $(20).$ The next step is to regularize the delta
distribution appearing in $(20)$ by a so-called strict delta net
$(\delta_{\varepsilon})_{\varepsilon\in(0,1]}\in C^{\infty}(\tilde{M})^{(0,1]}$,
sometimes called a model delta net, i.e. a net that has to meet the
following conditions\footnote{Note that, to simplify notations, condition $a)$ is often replaced
by an alternative condition $a')$ which requires that $supp(\delta_{\varepsilon})\subseteq[-\varepsilon,\varepsilon]\;\forall\varepsilon\in(0,1).$ }

\begin{align*}
a) & \;supp(\delta_{\varepsilon})\rightarrow\{0\}\quad(\varepsilon\rightarrow0)\\
b) & \;\int\delta_{\varepsilon}(x)dx\rightarrow1\quad(\varepsilon\rightarrow0)\:and\\
c) & \;\exists\eta>0\:\exists C\geq0:\int\vert\delta_{\varepsilon}(x)\vert dx\leq C\:\forall\varepsilon\in(0,\eta).
\end{align*}
More specifically, based on the observation that any such net converges
to a delta distribution as $\varepsilon\rightarrow0$, the regularized
profile function $f_{\varepsilon}=f_{\varepsilon}(u,v,\theta,\phi)$
of the geometry shall be chosen in such a way that

\begin{equation}
f_{\varepsilon}\equiv f_{0}\delta_{\varepsilon},
\end{equation}
where $f_{0}=f_{0}(v,\theta,\phi)$ and it shall be assumed that $(\delta_{\varepsilon})_{\varepsilon}$
is a function of $u$ only.

In somewhat sloppy notation, one can then write

\begin{equation}
\tilde{g}_{ab}^{\varepsilon}\equiv g_{ab}+f_{\varepsilon}l_{a}l_{b},
\end{equation}
and

\begin{equation}
\tilde{g}_{\varepsilon}^{ab}\equiv g^{ab}-f_{\varepsilon}l^{a}l^{b}.
\end{equation}
Note that the same idea was used in \cite{balasin1997geodesics,kunzinger1999rigorous,steinbauer1998geodesics}
to solve the geodesic and the geodesic deviation equations for impulsive
pp-wave spacetimes. 

Next, in order to be able to perform a variation of the Einstein-Hilbert
action for distributional metrics later on, the validity of $(14)$
shall be required. In addition, it shall be assumed that there holds

\begin{equation}
(l\nabla)f_{\varepsilon}=0
\end{equation}
and 
\begin{equation}
\tilde{R}_{ab}^{\varepsilon}l^{a}l^{b}=0,
\end{equation}
which can hold if and only if $(l\nabla)f_{0}=0$ and $R_{\;b}^{a}l_{a}l^{b}=0$
holds locally on $\mathcal{X}$ as well. 

Based on these assumptions, one comes to the conclusion that condtions
$(10-12)$ are met and thus relation $(17)$ is valid, which implies
that the defomed distributional Ricci and Einstein tensors are linear
in the generalized profile function $(f_{\varepsilon})_{\varepsilon}$.
In addition, based on the validity of $(m\nabla)l^{a}\propto l^{a}$
and $(l\nabla)(m\nabla)f_{0}=(m\nabla)(l\nabla)f_{\text{0}}=0$, it
is found that relation $(18)$ and thus relation $(19)$ applies in
this context as well, so that it can be concluded that also the deformed
distributional Riemann tensor is linear in the generalized profile
function $(f_{\varepsilon})_{\varepsilon}$. However, this ensures
that by construction the limit $\varepsilon\rightarrow0$ yields reasonable
singular expressions and that, in principle, 'standard' linear distribution
theory could be used in order to solve the field equations of the
theory. 

This shall turn out to be of great importance for the variation of
the Einstein-Hilbert action in the following. 

As is well known, in the standard smooth case, one uses the fact that
the total action of the system consists of two parts; a pure geometric
and - in the case that the gravitational field is coupled to a material
source - an additional matter part, so that there holds
\begin{equation}
S\equiv S_{\mathcal{G}}+S_{\mathcal{M}}.
\end{equation}
In the given generalized Kerr-Schild context, one has $S\equiv S[\tilde{g}]$,
$S_{\mathcal{G}}\equiv S_{\mathcal{G}}[\tilde{g}]\equiv\frac{1}{16\pi}\int\limits _{\tilde{M}}\tilde{R}\omega$
and $S_{\mathcal{M}}\equiv S_{\mathcal{M}}[\tilde{g}]\equiv\int\limits _{\tilde{M}}\tilde{T}\omega$,
where $\tilde{R}\equiv\tilde{R}_{a}^{a}\equiv\tilde{R}_{ab}\tilde{g}^{ab}$
is the scalar curvature, $\tilde{T}\equiv\tilde{T}_{a}^{a}\equiv\tilde{T}_{ab}\tilde{g}^{ab}$
is the scalar scalar energy-momentum density and $\omega\equiv\omega_{g}=\tilde{\omega}_{\tilde{g}}\equiv\tilde{\omega}$
is the scalar volume form being defined with respect to the determinant
$\tilde{g}=g$ of the Kerr-Schild metric $\tilde{g}_{ab}$ of spacetime
\cite{kramer1980exact}. Here, as usual, a variation of both parts
of the action yields (up to an irrelevant total divergence term)
\begin{equation}
\delta S\equiv\frac{1}{8\pi}\int\limits _{\tilde{M}}\tilde{G}_{ab}\delta\tilde{g}^{ab}\omega+\int\limits _{\tilde{M}}\tilde{T}_{ab}\delta\tilde{g}^{ab}\omega.
\end{equation}
By requiring then that $\delta S\overset{!}{=}0$, the field equations
of the theory result.

In the singular case, on the other hand, the situation is similar,
but not identical. Here again the total action consists of a purely
geometric and a matter part, but in the sense of distributions, which
means that the total action is associated with these parts in the
following sense

\begin{equation}
S\thickapprox S_{\mathcal{G}}+S_{\mathcal{M}},
\end{equation}
where $S_{\mathcal{G}}\equiv\frac{1}{8\pi}\underset{\varepsilon\rightarrow0}{\lim}\langle\int\limits _{\tilde{M}}\tilde{R}^{\varepsilon}\omega,\nu\rangle$
and $S_{\mathcal{M}}\equiv\underset{\varepsilon\rightarrow0}{\lim}\langle\int\limits _{\tilde{M}}\tilde{T}^{\varepsilon}\omega,\nu\rangle$
are defined with respect to the Kerr-Schild deformed generalized Ricci
and Laue scalars $(\tilde{R}^{\varepsilon})_{\varepsilon}$ and $(\tilde{T}^{\varepsilon})_{\varepsilon}$
(which can be calculated from metric $(31)$) and compactly supported
one-density $\nu$ on $\tilde{M}.$ Using these definitions, a combined
variation of both the geometric and the stress-energy part yields
the result 

\begin{equation}
\delta S\thickapprox\delta S_{\mathcal{G}}+\delta S_{\mathcal{M}},
\end{equation}
which may be re-written in the form

\begin{equation}
\delta S\equiv\underset{\varepsilon\rightarrow0}{\lim}\left\{ \frac{1}{8\pi}\langle\int\limits _{\tilde{M}}\left(\tilde{G}_{ab}^{\varepsilon}\delta\tilde{g}_{\varepsilon}^{ab}+\delta\tilde{R}_{ab}^{\varepsilon}\tilde{g}_{\varepsilon}^{ab}\right)\omega,\nu\rangle+\langle\int\limits _{\tilde{M}}\tilde{T}_{ab}^{\varepsilon}\delta\tilde{g}_{\varepsilon}^{ab}\omega,\nu\rangle\right\} ,
\end{equation}
where the second second part of the first integral $\langle\int\limits _{\tilde{M}}\delta\tilde{R}_{ab}^{\varepsilon}\tilde{g}_{\varepsilon}^{ab}\omega,\nu\rangle$
leads to a total divergence and thus to a boundary term, which, however,
shall be assumed to be zero for the sake of simplicity. 

Due to the fact that nets of generalized functions rather than distributions
are considered in this context, the given variation of the action
may turn out to be reasonable in the sense that the objects considered
are mathematically well-defined. This is not least because $\omega_{\varepsilon}\equiv\omega_{\tilde{g}_{\varepsilon}}=\omega_{g}\equiv\omega$
applies for the volume form of a generalized Kerr-Schild spacetime,
so that there is no dependence on the regularization in this case.
And also the very dangerous looking terms $\tilde{G}_{ab}^{\varepsilon}\delta\tilde{g}_{\varepsilon}^{ab}$
and $\tilde{T}_{ab}^{\varepsilon}\delta\tilde{g}_{\varepsilon}^{ab}$
turn out to be harmless in the final analysis, as $\tilde{G}_{ab}^{\varepsilon}\delta\tilde{g}_{\varepsilon}^{ab}\equiv\tilde{G}_{ab}^{\varepsilon}\delta g^{ab}$
and $\tilde{T}_{ab}^{\varepsilon}\delta\tilde{g}_{\varepsilon}^{ab}\equiv\tilde{T}_{ab}^{\varepsilon}\delta g^{ab}$
applies to them if the validity of $\tilde{G}_{ab}^{\varepsilon}\delta l^{a}l^{b}=\tilde{T}_{ab}^{\varepsilon}\delta l^{a}l^{b}=0$
is required in the present context, which is what is to be done in
the following. However, this implies that all dangerous, ill-defined
terms in $(39)$ turn out to be zero under the given cricumstances.

Hence, given this setting, it can now finally be required - in accordance
with the principle of stationary action - that

\begin{equation}
\delta S\thickapprox0.
\end{equation}
For Kerr-Schild spacetimes with geometric properties discussed above,
this variational principle then implies the validity of the distributional
field equations
\begin{equation}
\tilde{G}_{ab}\thickapprox8\pi\tilde{T}_{ab},
\end{equation}
where $\tilde{G}_{ab}$ and $\tilde{T}_{ab}$ are the embeddings of
the Einstein and energy-momentum tensors into the Colombeau algebra.
In this context, it may be noted that one is in principle free to
choose any (suitable) regularization of the action. However, to fix
a specific regularization and therefore to guarantee that the regularized
Einstein-Hilbert action is diffeomorphism invariant in the same way
as in the smooth case, one may decide to work not with the special,
but with the full diffeomorphism invariant Colombeau algebra treated
in \cite{grosser2012global,grosser2002global}. 

Examples for distributional spacetimes, in relation to which a variational
principle of this kind can actually be formulated, shall be discussed
in the next and final section of this work.

\section{Distributional Field Equations and Curvature}

The variational principle discussed in the previous section does not
apply to distributional Kerr-Schild spacetimes, but only to selected
ones, whose properties are very close to those discussed in section
one. Therefore, an overview of specific models shall now be given,
in relation to which said variation can be performed, whereupon the
main focus will be placed on setting up the field equations 'the right
way' via using Colombeau methods discussed in the previous section.
On the basis of the results to be determined, the consistency of the
method developed in the previous section for varying the regularized
total action of general relativity (gravity plus matter) is demonstrated.

A class of spacetimes to which these methods can be applied is the
class of impulsive plane fronted gravitational waves with parallel
rays or, for short, impulsive pp-waves, which is a class of singular
spacetimes where the curvature is concentrated on a null hypersurface.
As first discovered by Penrose \cite{penrose1972geometry}, these
spacetimes arise in the so-called impulsive limit of so-called sandwich
pp-waves for infinitesimal time intervals, or, as he further noted,
as a byproduct of the so-called 'scissors-and-paste' procedure, often
alternatively referred to as 'cut-and-paste' procedure in the literature. 

The line element of spacetimes belonging to this class can be written
in the form

\begin{equation}
ds^{2}=fdu^{2}-2dvdu+dy^{2}+dz^{2},
\end{equation}
where $f=f(v,u,y,z)$ is the so-called profile function of the geometry.
More precisely, the main characteristic of representatives of this
class is that their profile function can be written in the form
\begin{equation}
f(u,y,z)=\delta(u)f_{0}(x,y),
\end{equation}
where $f_{0}(x,y)$ is the reduced profile function and $\delta(u)$
is Dirac's delta distribution. Consequently, as can be concluded from
the observation that the metric related to line element $(45)$ can
be written in the form
\begin{equation}
g_{ab}=\eta_{ab}+f_{0}\delta l_{a}l_{b},
\end{equation}
the given spacetime is flat everywhere except for the null hyperplane
$u=0$, where the delta-like impulse is located. Representatives of
the given class of geometries therefore belong to the superordinate
Kerr-Schild class of spacetimes.

As first discovered by Ehlers and Kundt \cite{ehlers1962gravitation,kramer1980exact,kundt1961plane},
the family of pp-wave spacetimes, however, also belongs to another
superordinate family of spacetimes at the same time, the so-called
Kundt family; a family of non-twisting, shear-free and non-expanding
geometries, whose line element is of the form

\begin{equation}
ds^{2}=Hdu^{2}-2dvdu+2W_{b}dx^{b}dv+q_{bc}dx^{b}dx^{c},
\end{equation}
where $H=H(v,u,x^{2},x^{3})$, $W_{b}=W_{b}(v,u,x^{2},x^{3})$ and
$q_{bc}=q_{bc}(v,x^{2},x^{3})$ applies by definition. 

The main characteristic of this class is that there always exists
a null vector field $l^{a}=\partial_{v}^{a}$, which is the generator
of a null foliation of spacetime in non-expanding null hypersurfaces,
so that it can be concluded that condition $(14)$ is always globally
met. In addition, $f$ can always be chosen in such a way for pp-wave
geometries that condition $(15)$ is met as well. Besides that, it
can be arranged that $(m\nabla)l^{a}\propto l^{a}$ and $(l\nabla)(m\nabla)f=(m\nabla)(l\nabla)f=0$
also holds for these spacetimes and the Ricci tensor $R_{ab}$ of
the geometry is always given in such a way that $R_{ab}\propto l_{a}l_{b}$,
so that it can be concluded that not only conditions $(10-12)$ are
met, but also relation $(18)$ turns out to be valid, which, however,
implies that the validity of $(17)$ and $(19)$.

The exact same applies in the case of impulsive pp-waves; however,
only in a distributional sense. This implies that not only the deformed
Ricci tensor with lowered or raised indices is linear in the profile
function, but also the deformed part of the Riemann curvature tensor.
In addition, using that there holds $\omega_{\varepsilon}\equiv\omega_{\tilde{g}_{\varepsilon}}=\omega_{g}\equiv\omega$
for the volume form of a generalized Kerr-Schild spacetime, it becomes
possible to set up a regularized gravitational action (matter plus
gravity) of the form $(37)$ and to repeat the steps discussed in
the previous section, which lead to distributional field equations
of the form $(41)$. 

Specific models in general relativity to which this approach can be
applied include the arguably most well-known model for an impulsive
pp-wave spacetime, the geometric model of Aichelburg and Sexl, whose
reduced profile function is given by the expression
\begin{equation}
f_{0}(y,z)=8p\ln\sqrt{y^{2}+z^{2}}.
\end{equation}
This spacetime, which \textcolor{black}{was originally found by the
authors by calculating the ultrarelativistic limit of a Lorentz boosted
Schwarzschild geometry in isotropic coordinates}, characterizes the
field of a point-like particle moving close to the speed of light.
It represents an exact solution to Einstein's equations, which in
the given case reduce to a single differential equation for the reduced
profile function of the form 
\begin{equation}
\Delta_{\mathbb{R}_{2}}f_{0}=-16\pi p\delta^{(2)},
\end{equation}
where $\delta^{(2)}\equiv\delta^{(2)}(y,z)$. 

A similar model, to which the ideas formulated in section two also
apply, was found by Lousto and Sanchez \cite{lousto1990curved}, who
calculated the ultrarelativistic limit of the boosted Reissner-Nordström
geometry. Assuming that $e=\gamma^{\frac{1}{2}}p_{e}$, they had to
require for this purpose the charge $e$ to vanish in the said limit.
Based on this constraint, they found a solution of Einstein's equations
of the form $(47)$ with reduced profile function
\begin{equation}
f_{0}(y,z)=8p\ln\sqrt{y^{2}+z^{2}}+\frac{3\pi p_{e}^{2}}{2\sqrt{y^{2}+z^{2}}}.
\end{equation}
The solution obtained again represents a pp-wave and is flat everywhere
except on the null plane $u=0.$ More precisely, as was confirmed
by Steinbauer in \cite{steinbauer1997ultrarelativistic} using a calculation
in $\mathcal{G}$, it was found that that all the components of the
electromagnetic field and all but one of the components of the energy-momentum
tensor are associated to zero. 

The ultrarelativistic limit of the Kerr metric has been calculated
by several authors \cite{balasin1995ultrarelativistic,barrabes2003lightlike,barrabes2004scattering,ferrari1990boosting,lousto1989gravitational,lousto1989ultrarelativistic}.
The same limit has also been applied to other non-flat backgrounds
and used for a number of different boosted sources, such as cosmolgical
constant sources, cosmic strings, domain walls and monopoles to obtain
ultrarelativistic impulsive pp-wave spacetimes \cite{hotta1993shock,lousto1990curved,lousto1991gravitational,podolsky1998boosted,podolsky1998impulsive,podolsky1999expanding},
which have been used to describe (quantum) scattering processes of
highly energetic particles \cite{lousto1989ultrarelativistic,lousto1992ultrarelativistic}.
Of course, the methods developed in section two of this paper work
for all these approaches in a similar way. 

In any case, taking into account the results obtained in section one
of this work, it becomes clear that there is another class of geometries
to which these methods can be applied, namely the class of gravitational
shock wave geometries in black hole and cosmological backgrounds. 

The most famous representatives of this class were all found on the
basis of Penrose's cut-and-paste procedure, one of the forerunners
of today's thin shall approaches \cite{mars1993geometry,senovilla2015double,sfetsos1995gravitational}.
In particular, the field of a spherical shock wave caused by a massless
particle moving at the speed of light along the horizon of a Schwarzschild
black hole was derived by Dray and \textquoteright t Hooft \cite{dray1985gravitational}
on the basis of Penrose's methods. Using exactly the same method,
Sfetsos calculated a similar geometry for the Reissner-Nordström case
\cite{sfetsos1995gravitational} and Lousto and Sanchez specified
a spherical shock wave for the Kottler alias Schwarzschild-de Sitter
case. 

Due to their similarity, all these approaches shall be discussed in
a single effort in the following. The reason why this is possible
is the following: Using the fact that the line element of any static
spherically symmetric spacetime can be brought into the form
\begin{equation}
-2A^{2}dUdV+r^{2}(d\theta^{2}+\sin^{2}\theta d\phi^{2}),
\end{equation}
\textcolor{black}{where $r=r(UV)$ and $A=A(r(UV))$ are implicit}
\textcolor{black}{functions }of $U$ and $V$, the line elements of
the Dray-'t Hooft, Sfetsos and Lousto-Sanchez shock wave geometries
written down as follows

\begin{equation}
ds^{2}=2A^{2}f_{0}\delta dU^{2}-2A^{2}dUdV+r^{2}(d\theta^{2}+\sin^{2}\theta d\phi^{2})
\end{equation}
where $\delta=\delta(U)$ is Dirac's delta distribution. Hence, it
can be concluded that the metrics corresponding to these line elements
belong individually to the generalized Kerr-Schild classes
\begin{equation}
\tilde{g}_{ab}=g_{ab}+2A^{2}f_{0}\delta l_{a}l_{b},
\end{equation}
of the Schwarzschild, Reissner-Nordström and Kottler backgrounds,
where in each case one has $l_{a}=g_{ab}l^{b}=-dU_{a}$. 

What all three cases additionally have in common is that the profile
functions of the respective shock wave geometries can be obtained
via solving Einstein's equations, whereupon in all three cases ill-defined
'delta-square' terms occurred in the course of the calculation, which,
from a mathematical point of view, made no sense at all and therefore
had to be ignored by the authors in each and every case. For this
reason, in particular, it was shown several years later by Balasin
on the basis of the generalized Kerr-Schild framework that the problematic
terms do not occur in Einstein's equations with mixed indices \cite{balasin2000generalized},
which allowed him to rigorously deduce said shock wave spacetimes
from their associated backgrounds\footnote{To be exact, Balasin did not demonstrate explicitly the validity of
his method for all these cases, but only for the geometry of Dray
and 't Hooft. However, his results are completely general and therefore,
of course, turn out to be valid for the cases mentioned above as well. }. Thus, as can straightforwardly be deduced from Balasin's results,
the mixed field equations of the generalized Kerr-Schild class lead
to a single differential equation for the reduced profile function
of the form 

\begin{equation}
(\Delta_{\mathbb{S}_{2}}-c)f_{0}=2\pi b\delta,
\end{equation}
where $\delta\equiv\delta(\cos\theta-1)$ is Dirac's delta distribution
and $b$ and $c$ are constants, whereas $c$ is given by $c=2r_{+}(\kappa-\Lambda r_{+})$
in the Schwarschild-de Sitter case, $c=2r_{+}\kappa$ in the Reissner-Nordström
case and by $c=1$ in the Schwarzschild case. 

The resulting equation can be solved by expanding the reduced profile
function on the left hand side and the delta function on the right
hand side simultaneously in Legendre polynomials. Using here the fact
that $\delta(x)=\overset{\infty}{\underset{l=0}{\sum}}(l+\frac{1}{2})P_{l}(x)$,
one obtains the solution
\begin{equation}
f_{0}(\theta)=-b\overset{\infty}{\underset{l=0}{\sum}}\frac{l+\frac{1}{2}}{l(l+1)+c}P_{l}(\cos\theta)
\end{equation}
by solving the corresponding eigenvalue problem. An integral expression
for this solution can then be found by considering the generating
function of the Legendre polynomials
\begin{equation}
\overset{\infty}{\underset{l=0}{\sum}}\frac{l+\frac{1}{2}}{l(l+1)+c}P_{l}(\cos\theta)e^{-sl}=\frac{1}{\sqrt{1-2\cos\theta e^{-s}+e^{-2s}}}
\end{equation}
in addition to the fact that 

\begin{equation}
\frac{l+\frac{1}{2}}{l(l+1)+\alpha^{2}+\frac{1}{4}}=\underset{0}{\overset{\infty}{\int}}e^{-s(l+\frac{1}{2})}\cos(\alpha s)ds.
\end{equation}
This yields a result of the form
\begin{equation}
f_{0}(\theta)=-\frac{b_{0}}{\sqrt{2}}\underset{0}{\overset{\infty}{\int}}\frac{\cos(\sqrt{c-\frac{1}{4}}s)}{\sqrt{\cosh s-\cos\theta}}ds,
\end{equation}
which for each individual value of\textcolor{black}{{} $r=r(UV)$ and
$A=A(r(UV))$ and} $c$ gives the precise form of the shock wave geometries
of Dray and 't Hooft, Sfetsos and Lousto and Sanchez.

The only problem with this idea of treating the subject is that it
still proves to be impossible to deduce the corresponding distributional
geometries from Einstein's equations with lowered and raised indices.
Besides that, one could come to the conclusion that it is still impossible
to set up a regularized action of the form $(37)$ for such spacetimes
and to obtain distributional field equations of the form $(41)$ from
its variation.

However, based on the methods developed in section two, it becomes
clear that this is in fact not true. Using once more the fact that
$\omega_{\varepsilon}\equiv\omega_{\tilde{g}_{\varepsilon}}=\omega_{g}\equiv\omega$
for the volume form of generalized Kerr-Schild spacetimes and that
condition $(14)$ is met and that also $(33)$ turns out to be valid
if it is required that $\partial_{V}A\vert_{U=0}=\partial_{V}r\vert_{U=0}=0$,
it can be checked that conditions $(12)$, $(14)$, $(15)$ and $(18)$
are met and therefore relations $(17)$ and $(19)$ turn out to be
valid in the given setting, which is not least due to the fact that
$R_{ab}l^{a}l^{b}=0$ holds true for the three different background
fields. 

Thus, it can be concluded that Balasin's ideas can be extended by
a more specific choice of the Kerr-Schild approach to the effect that
said problematic 'delta-square' terms never occur in the field equations
of the theory and even the deformed Riemannian curvature tensor (if
calculated appropriately) does not contain any ill-defined non-linear
delta-terms. As a result, it becomes possible to set up a regularized
gravitational action (matter plus gravity) of the form $(37)$ and
to repeat the steps discussed in the previous section, which lead
to distributional field equations of the form $(41)$. 

Finally, note that quite recently Balasin's results were used as a
starting point for calculating the field of a gravitational shock
wave caused by a massless particle moving at the speed of light along
the exterior event horizon of a Kerr-Newman black hole \cite{Huber2019ultpar}.
Using the same geometrical assumptions as in the present work, a much
more general class of spacetimes was deduced in this context, to which
the variation principle for distributional Kerr-Schild metrics developed
in section two can also be applied without further ado. 

\section*{Discussion}

In the present work, a subclass of the generalized Kerr-Schild spacetime
class was specified with respect to which the variational principle
of general relativity can be generalized - on the basis of Colombeau's
theory of generalized functions - to singular situations, in which
one would usually not expect the principle of stationary action to
be valid. Considering specific metrics in this context, which contain
a generalized delta (profile) function converging to a delta distribution
in an appropriate limit, it was shown how said variational principle
leads to mathematically exact distributional field equations. As further
shown for the cases of impulsive pp-wave spacetimes and various gravitational
shock wave geometries in black holes and cosmological backgrounds,
the solutions of these distributional field equations fit perfectly
into the physically expected picture, as they contain no undefined
'delta-square' terms or the like, proving yet again why Colombeau's
theory of generalized functions offers not only a solid mathematical
framework, but an indispensable machinery for accommodating calculations
involving distributional metrics in general relativity as well.
\begin{description}
\item [{Acknowledgements:}]~
\end{description}
I want to thank Roland Steinbauer for a very interesting and insightful
discussion on the subject.

\bibliographystyle{plain}
\addcontentsline{toc}{section}{\refname}\bibliography{1C__Arbeiten_litcol}

\begin{thebibliography}{10}

\bibitem{aichelburg1971gravitational}
{Peter}~{C} Aichelburg and {Roman}~{U} {Sexl}.
\newblock On the gravitational field of a massless particle.
\newblock {\em General {Relativity} and {Gravitation}}, 2(4):303--312, 1971.

\bibitem{alonso1987generalized}
{Rodrigo} Alonso and {Nelson} {Zamorano}.
\newblock Generalized {Kerr}-{Schild} metric for a massless particle on the
  {Reissner}-{Nordstr{\"o}m} horizon.
\newblock {\em Physical {Review} {D}}, 35(6):1798, 1987.

\bibitem{balasin1997geodesics}
{Herbert} Balasin.
\newblock Geodesics for impulsive gravitational waves and the multiplication of
  distributions.
\newblock {\em Classical and {Quantum} {Gravity}}, 14(2):455, 1997.

\bibitem{balasin2000generalized}
{Herbert} Balasin.
\newblock Generalized {Kerr}-{Schild} metrics and the gravitational field of a
  massless particle on the horizon.
\newblock {\em Classical and {Quantum} {Gravity}}, 17(9):1913, 2000.

\bibitem{balasin1995ultrarelativistic}
{Herbert} Balasin and {Herbert} {Nachbagauer}.
\newblock The ultrarelativistic {Kerr} geometry and its energy-momentum tensor.
\newblock {\em Classical and {Quantum} {Gravity}}, 12(3):707, 1995.

\bibitem{barrabes2003lightlike}
{C} Barrabes and {P}{A} {Hogan}.
\newblock Lightlike boost of the {Kerr} gravitational field.
\newblock {\em Physical {Review} {D}}, 67(8):084028, 2003.

\bibitem{barrabes2004scattering}
C~Barrab{\`e}s and PA~Hogan.
\newblock Scattering of high-speed particles in the kerr gravitational field.
\newblock {\em Physical {Review} {D}}, 70(10):107502, 2004.

\bibitem{colombeau2000new}
Jean~Fran{\c{c}}ois Colombeau.
\newblock {\em New generalized functions and multiplication of distributions}.
\newblock Elsevier, 2000.

\bibitem{colombeau2011elementary}
Jean~Fran{\c{c}}ois Colombeau.
\newblock {\em Elementary introduction to new generalized functions}.
\newblock Elsevier, 2011.

\bibitem{dray1985gravitational}
{Tevian} Dray and {Gerard} 't~{Hooft}.
\newblock The gravitational shock wave of a massless particle.
\newblock {\em Nuclear {Physics} {B}}, 253:173--188, 1985.

\bibitem{ehlers1962gravitation}
{J{\"u}rgen} Ehlers and {Wolfgang} {Kundt}.
\newblock Gravitation: An {Introduction} to {Current} {Research} ed {L}
  {Witten}, 1962.

\bibitem{ferrari1990boosting}
{Valeria} Ferrari and {Paolo} {Pendenza}.
\newblock Boosting the {Kerr} metric.
\newblock {\em General Relativity and Gravitation}, 22(10):1105--1117, 1990.

\bibitem{friedrich1999rigidity}
{Helmut} {Friedrich}, {Istvan} {Racz}, and {Robert}~{M} {Wald}.
\newblock On the rigidity theorem for spacetimes with a stationary event
  horizon or a compact cauchy horizon.
\newblock {\em {Communications} in mathematical physics}, 204(3):691--707,
  1999.

\bibitem{grosser2001geometric}
M.~Grosser, M.~Kunzinger, M.~Oberguggenberger, and R.~Steinbauer.
\newblock {\em Geometric theory of generalized functions with applications to
  general relativity}, volume 537.
\newblock Springer Science \& Business Media, 2001.

\bibitem{grosser2012global}
{M} {Grosser}, {M} {Kunzinger}, {R} {Steinbauer}, and {J}{A} {Vickers}.
\newblock A {Global} {Theory} of {Algebras} of {Generalized} {Functions} ii:
  tensor distributions.
\newblock {\em New York Journal of Mathematics}, 18:139--199, 2012.

\bibitem{grosser2001foundations}
{Michael} Grosser, {Eva} {Farkas}, {Michael} {Kunzinger}, and {Roland}
  {Steinbauer}.
\newblock {\em On the foundations of nonlinear generalized functions I and II}.
\newblock Number 153. American {Mathematical} {Soc.}, 2001.

\bibitem{grosser2002global}
{Michael} Grosser, {Michael} {Kunzinger}, {Roland} {Steinbauer}, and
  {James}~{A} {Vickers}.
\newblock A global theory of algebras of generalized functions.
\newblock {\em Advances in {Mathematics}}, 166(1):50--72, 2002.

\bibitem{hayward1993dual}
{Sean}~{A} Hayward.
\newblock Dual-null dynamics of the {Einstein} field.
\newblock {\em Classical and {Quantum} {Gravity}}, 10(4):779, 1993.

\bibitem{hotta1993shock}
{M} Hotta and {M} {Tanaka}.
\newblock Shock-wave geometry with nonvanishing cosmological constant.
\newblock {\em Classical and {Quantum} {Gravity}}, 10(2):307, 1993.

\bibitem{Huber2019ultpar}
{Albert} Huber.
\newblock The gravitational {Field} of a massless {Particle} on the {Horizon}
  of a stationary {Black} {Hole}.
\newblock {\em arxiv preprint, arXiv:1911.02299}.

\bibitem{Huber2019foliation}
{Albert} Huber.
\newblock Null foliations and the geometry of black hole horizons.
\newblock {\em arxiv preprint, arXiv:1908.08739}.

\bibitem{kramer1980exact}
{D} Kramer, {Hans} {Stephani}, {M} {Mac}{Callum}, and {E} {Herlt}.
\newblock Exact solutions of {Einsteins} field equations.
\newblock {\em Berlin}, 1980.

\bibitem{kundt1961plane}
{Wolfgang} Kundt.
\newblock The plane-fronted gravitational waves.
\newblock {\em Zeitschrift f{\"u}r {Physik}}, 163(1):77--86, 1961.

\bibitem{kunzinger1999rigorous}
{Michael} Kunzinger and {Roland} {Steinbauer}.
\newblock A rigorous solution concept for geodesic and geodesic deviation
  equations in impulsive gravitational waves.
\newblock {\em Journal of {Mathematical} {Physics}}, 40(3):1479--1489, 1999.

\bibitem{kunzinger2002foundations}
{Michael} {Kunzinger} and {Roland} {Steinbauer}.
\newblock Foundations of a nonlinear distributional geometry.
\newblock {\em Acta {Applicandae} {Mathematica}}, 71(2):179--206, 2002.

\bibitem{kunzinger2002generalized}
{Michael} Kunzinger and {Roland} {Steinbauer}.
\newblock Generalized pseudo-{Riemannian} geometry.
\newblock {\em Transactions of the {American} {Mathematical} {Society}},
  354(10):4179--4199, 2002.

\bibitem{kunzinger2009sheaves}
Michael Kunzinger, Roland Steinbauer, and James Vickers.
\newblock Sheaves of nonlinear generalized functions and manifold-valued
  distributions.
\newblock {\em Transactions of the American Mathematical Society},
  361(10):5177--5192, 2009.

\bibitem{kunzinger2003intrinsic}
{Michael} Kunzinger, {Roland} {Steinbauer}, and {James}~{A} {Vickers}.
\newblock Intrinsic characterization of manifold-valued generalized functions.
\newblock {\em Proceedings of the {London} {Mathematical} {Society}},
  87(2):451--470, 2003.

\bibitem{lousto1989gravitational}
{C}{O} Lousto and {N} {S{\'a}nchez}.
\newblock Gravitational shock waves of ultra-high energetic particles on curved
  spacetimes.
\newblock {\em Physics {Letters} {B}}, 220(1-2):55--60, 1989.

\bibitem{lousto1989ultrarelativistic}
{C}{O} Lousto and {N} {S{\'a}nchez}.
\newblock The ultrarelativistic limit of the {Kerr}-{Newman} geometry and
  particle scattering at the {Planck} scale.
\newblock {\em Physics {Letters} {B}}, 232(4):462--466, 1989.

\bibitem{lousto1990curved}
{C}{O} Lousto and {N} {S{\'a}nchez}.
\newblock The curved shock wave space-time of ultrarelativistic charged
  particles and their scattering.
\newblock {\em International {Journal} of {Modern} {Physics} {A}},
  5(05):915--938, 1990.

\bibitem{lousto1991gravitational}
{C}{O} Lousto and {N} {S{\'a}nchez}.
\newblock Gravitational shock waves generated by extended sources:
  {Ultrarelativistic} cosmic strings, monopoles and domain walls.
\newblock {\em Nuclear {Physics} {B}}, 355(1):231--249, 1991.

\bibitem{lousto1992ultrarelativistic}
{C}{O} Lousto and {N} {S{\'a}nchez}.
\newblock The ultrarelativistic limit of the boosted {Kerr}-{Newman} geometry
  and the scattering of spin-1/2 particles.
\newblock {\em Nuclear {Physics} {B}}, 383(1-2):377--394, 1992.

\bibitem{mars1993geometry}
{Marc} Mars and {Jose}~{M}{M} {Senovilla}.
\newblock Geometry of general hypersurfaces in spacetime: junction conditions.
\newblock {\em Classical and {Quantum} {Gravity}}, 10(9):1865, 1993.

\bibitem{moncrief1983symmetries}
{Vincent} Moncrief and {James} {Isenberg}.
\newblock Symmetries of cosmological {Cauchy} horizons.
\newblock {\em Communications in {Mathematical} {Physics}}, 89(3):387--413,
  1983.

\bibitem{nigsch2015new}
{E}{A} Nigsch.
\newblock A new approach to diffeomorphism invariant algebras of generalized
  functions.
\newblock {\em Proceedings of the [Edinburgh}.

\bibitem{nigsch2016full}
{Eduard}~A Nigsch and {Michael} {Grosser}.
\newblock Full and special {Colombeau} algebras.
\newblock {\em arXiv preprint arXiv:1611.06061}, 2016.

\bibitem{penrose1972geometry}
{Roger} Penrose.
\newblock The geometry of impulsive gravitational waves.
\newblock {\em General {Relativity}, {Papers} in {Honour} of {J}{L} {Synge}},
  pages 101--115, 1972.

\bibitem{podolsky1998boosted}
{Ji{\v{r}}{\'\i}} Podolsk{\`y} and {Jerry}~{B} {Griffiths}.
\newblock Boosted static multipole particles as sources of impulsive
  gravitational waves.
\newblock {\em Physical {Review} {D}}, 58(12):124024, 1998.

\bibitem{podolsky1998impulsive}
{Ji{\v{r}}{\'\i}} Podolsk{\`y} and {Jerry}~{B} {Griffiths}.
\newblock Impulsive waves in de {Sitter} and anti-de {Sitter} spacetimes
  generated by null particles with an arbitrary multipole structure.
\newblock {\em Classical and {Quantum} {Gravity}}, 15(2):453, 1998.

\bibitem{podolsky1999expanding}
{Ji{\v{r}}{\'\i}} Podolsk{\`y} and {Jerry}~{B} {Griffiths}.
\newblock Expanding impulsive gravitational waves.
\newblock {\em Classical and {Quantum} {Gravity}}, 16(9):2937, 1999.

\bibitem{schwartz1954limpossibilite}
Laurent Schwartz.
\newblock Sur l'impossibilite de la multiplication des distributions.
\newblock {\em Comptes {Rendus} {Hebdomadaires} des {Seances} de {L}'
  {Academie} des {Sciences}}, 239(15):847--848, 1954.

\bibitem{senovilla2015double}
{Jos{\'e}}~{M}{M} Senovilla.
\newblock Double layers in gravity theories.
\newblock In {\em Journal of {Physics}: {Conference} {Series}}, volume 600,
  page 012004. I{O}{P} {Publishing}, 2015.

\bibitem{sfetsos1995gravitational}
{Konstadinos} Sfetsos.
\newblock On gravitational shock waves in curved spacetimes.
\newblock {\em Nuclear {Physics} {B}}, 436(3):721--745, 1995.

\bibitem{steinbauer2010geometric}
R~{Steinbauer}.
\newblock A geometric approach to full {Colombeau} algebras.
\newblock {\em Banach {Center} {Publications}}, 88:267--272, 2010.

\bibitem{steinbauer1997ultrarelativistic}
{Roland} Steinbauer.
\newblock The ultrarelativistic {Reissner}--{Nordstro/m} field in the
  {Colombeau} algebra.
\newblock {\em Journal of {Mathematical} {Physics}}, 38(3):1614--1622, 1997.

\bibitem{steinbauer1998geodesics}
{Roland} Steinbauer.
\newblock Geodesics and geodesic deviation for impulsive gravitational waves.
\newblock {\em Journal of {Mathematical} {Physics}}, 39(4):2201--2212, 1998.

\end{thebibliography}

\end{document}